\documentclass[twocolumn]{aastex631}
\usepackage{amsmath}	
\usepackage{amssymb}	
\usepackage{hyperref}
\usepackage{newtxtext,newtxmath}
\usepackage{longtable}
\usepackage{tabularx}
\usepackage{float}
\usepackage{comment}
\usepackage{ragged2e}
\renewcommand{\baselinestretch}{1.10}
\usepackage{todonotes}

\received{October 28, 2025}
\revised{February 26, 2026}
\accepted{March 20, 2026}

\shorttitle{SED modeling of FSRQs}
\shortauthors{Bora et al.}

\begin{document}

\title[Jet power FSRQs]{Jet Power Estimates of FSRQs PKS 1441+25 and Ton 599 from Broadband SED Modeling}

\correspondingauthor{Hritwik Bora}
\email{hritwikbora@gmail.com}

\author[0000-0002-1520-057X]{Hritwik Bora}
\affiliation{Department of Physics, Tezpur University, Tezpur-784028, India}

\author[0000-0002-7609-2779]{Ranjeev Misra}
\affiliation{Inter-University Centre for Astronomy and Astrophysics, Post Bag 4, Ganeshkhind, Pune, Maharashtra - 411007}


\author[0000-0002-6941-7002]{Rukaiya Khatoon}
\affiliation{Centre for Space Research, North-West University, South Africa}

\author[0000-0003-1715-0200]{Rupjyoti Gogoi}
\affiliation{Department of Physics, Tezpur University, Tezpur-784028, India}



\begin{abstract}

Flat-Spectrum Radio Quasars (FSRQs) are among the most energetic and powerful active galactic nuclei, often exhibiting jet powers comparable to or exceeding the Eddington luminosity. In this work, we performed broadband spectral energy distribution (SED) modeling of two FSRQs PKS 1441+25 and Ton 599, using \textit{Swift}-XRT/UVOT, \textit{NuSTAR}, \textit{Fermi}-LAT and VERITAS observations during 2015 and 2021, respectively. We considered four particle distribution models: a broken power law, a log-parabola, and two energy-dependent models in which either the diffusion or acceleration timescale depends on energy. Our results show that the jet power estimates derived from models with intrinsic curvature, such as the log-parabola and energy-dependent models, are of the same order as those obtained with a broken power-law distribution. This contrasts with the case of High Synchrotron Peaked Blazars (HBLs), where the power estimates can differ by nearly two orders of magnitude between models. We attribute this difference to the lower electron break energies typically observed in FSRQs. Consequently, our findings suggest that, unlike in HBLs, the estimated jet powers in FSRQs are relatively insensitive to the assumed particle energy distribution, reflecting the dominance of external Compton processes and weaker dependence on spectral curvature.

\end{abstract}




\section{Introduction} \label{sec:intro}

Blazars are a subclass of active galactic nuclei (AGNs) characterized by relativistic jets pointed nearly along the observer’s line of sight \citep{urry1995unified}. These objects harbor a supermassive black hole (SMBH) at their centers and emit non-thermal radiation across a broad range of wavelengths, from radio to TeV gamma rays. Their spectral energy distributions (SEDs) typically exhibit a double-hump structure, along with properties such as rapid variability and radio-loudness \citep{urry1998multiwavelength,massaro2004log}. Blazars are divided into two primary categories based on their emission line features: Flat Spectrum Radio Quasars (FSRQs) and BL Lacs. (FSRQs), which show prominent emission lines, and BL Lacertae objects (BL Lacs), which exhibit weak or no emission lines \citep{1995PASP..107..803U,antonucci1993unified}. Based on the emission humps, blazars are further classified into different sub classes. 

FSRQs have their synchrotron peak frequency (low-energy hump) at $\sim$ 10$^{13.3}$ Hz, while for BL Lac objects, the peak frequencies are within 10$^{14}$-10$^{17}$ Hz for Low - Synchrotron peaked (LSP/LBL), Intermediate peaked BL Lacs (ISP/IBL) and high synchrotron peaked (HSP/HBL) BL Lacs \citep{2010ApJ...716...30A}. For FSRQs and LSP/LBL, the synchrotron peak arises in the far-Infrared (IR) or in the IR band of electromagnetic (EM) spectrum, while for ISP/IBL and HSP/HBL sources the synchrotron peak arises between the optical/Ultraviolet (UV) and  X-ray band respectively \citep{ghisellini1997optical, 1998MNRAS.299..433F}. 

The origin of the high-energy hump is still unresolved. In the leptonic scenario, the high-energy emission is attributed to inverse Compton (IC) scattering. In case of FSRQs, the seed photons might be external photons outside the jet (i.e. external Compton (EC)), or the synchrotron photons which are upscattered to higher energies (i.e. Synchrotron self Compton (SSC) in case of BL Lacs \citep{1985ApJ...298..114M, 1992A&A...256L..27D, 1993ApJ...416..458D, 1994ApJ...421..153S, 2014MNRAS.438..779G, 2018MNRAS.473.2639G}.  

One of the key challenges in understanding these systems is identifying the energy-generation process responsible for the observed high jet powers, as inferred from broadband spectral fitting. \cite{2003ApJ...593..667M} analyzed the SEDs of 11 FSRQs, and found the jet powers to be in the range of $(10^{46}-10^{48})$ erg s$^{-1}$. The upper limits on the powers of the sources were derived assuming a minimum electron Lorentz factor, $\gamma_\text{min} = 1$; increasing $\gamma_\text{min}$ would result in lower jet power estimates. In another study, \cite{Chen_2018}, using broadband SED modeling, estimated the jet properties of $\gamma-$ray loud AGNs for a three-parameter log-parabolic function. The reported mean value of the jet powers in FSRQs is in the order of $\sim 10^{45}$ erg s$^{-1}$. 

Several earlier studies have shown that the total jet power is larger than the disk luminosity for both FSRQs and BL lacs \citep{Ghisellini_2014,10.1093/mnras/stw1730,Chen_2018}. Even in various instances, it was found that, for several $\gamma$-ray loud and quiet blazars, the total jet powers average $\sim 10^{47}$ erg s$^{-1}$ with a few sources exceeding $10^{48}$ erg s$^{-1}$ \citep[e.g][]{Paliya_2017}. These estimates assume that the kinetic energy of cold protons dominates the jet power, under the condition that the proton number density matches that of non-thermal electrons. However, the total power requirement could be reduced by invoking a significant presence of electron–positron pairs. Despite this, the pair fraction is probably minimal because of Compton drag effects and the lack of associated spectral signatures \citep[e.g.][]{Sikora_2000, 10.1111/j.1365-2966.2007.12758.x, 10.1093/mnras/stw107}. If the hadronic picture is considered instead of the leptonic one, the power requirement becomes even greater, exceeding $10^{49}$ erg s$^{-1}$ \citep[e.g.][]{10.1093/mnrasl/slv039, Abe_2023}.  

Typically, the electron energy distribution is assumed to follow a broken power-law, with the number of electrons depending on the lowest Lorentz factor, $\gamma_\text{min}$, which is often taken to be around unity or in the tens. However, X-ray and $\gamma$-ray observations provide evidence that the observed spectrum exhibits significant curvature, deviating from this power-law behavior \citep{massaro2004log, Tanihata_2004, tramacere}. This suggests that the electron energy distribution itself may be curved, such as in a log-parabolic (LP) distribution \citep{massaro2004log}. Several other physically motivated models have also been proposed, including those where the high-energy cutoff results from radiative losses (the $\gamma$-max models), or where the curvature arises due to energy-dependent acceleration (EDA) or energy-dependent diffusion (EDD) \citep{Sinha_2017,10.1093/mnras/sty2003,Hota_2021,Khatoon_2022}. Although these different interpretations are often spectrally degenerate, they could potentially be distinguished by analyzing the physical significance of observed correlations between best-fit spectral parameters across various models \citep{Hota_2021,Khatoon_2022}.

In a recent study, \cite{10.1093/mnras/stae706} demonstrated that for the BL Lac source Mkn 501, adopting larger values of the minimum Lorentz factor ($\gamma_\text{min}$) significantly reduces the estimated jet power when assuming a broken power-law (BPL) electron distribution. In contrast, models incorporating intrinsic curvature, such as the log-parabola (LP) and those with energy-dependent diffusion or acceleration timescales, yielded jet powers of the order of $\sim 10^{43}$ erg s$^{-1}$, substantially lower than those derived from the BPL model. Similarly, \cite{tantry2024probingbroadbandspectralenergy} analyzed Mkn 501 during a low-activity state and found that the LP and EDA models resulted in lower jet power estimates than the broken power-law. These results suggest that, for HBLs, adopting curved particle distributions can considerably reduce the inferred jet powers, which may amount to only $\sim 10\%$ of the Eddington luminosity-implying that the jets could be powered primarily by accretion processes.

Given these findings, it is important to test whether similar trends hold for Flat Spectrum Radio Quasars (FSRQs), which exhibit different radiative environments and generally lower electron break energies than HBLs. Therefore, this work aims to investigate how alternative electron energy distributions beyond the standard broken power-law affect the jet power estimates in FSRQs. To this end, we model the broadband spectral energy distributions (SEDs) of two FSRQs, PKS 1441+25 and Ton 599, using multiwavelength data from Swift-XRT/UVOT, NuSTAR, and Fermi-LAT, focusing on periods of enhanced activity in April 2015 and June 2021, respectively.

PKS 1441+25 is an FSRQ located at redshift, z $\approx$ 0.939, with RA =  220$^\circ$.9 and DEC =  25$^\circ$.02. In April 2015, MAGIC and VERITAS collaborations detected VHE $\gamma$-rays from this source, at energies exceeding 80 GeV and reaching up to 250 GeV \citep{2015ATel.7416....1M, 2015ATel.7433....1M, 2017MNRAS.470.2861S}. As reported by \cite{2015ApJ...815L..22A}, the UV/X-ray emission for PKS 1441+25 is well described by synchrotron emission from a power law distribution. As expected for FSRQs, the high-energy hump in PKS 1441+25 arises from the EC process \citep{Ahnen_2015}. 

Ton 599 is an FSRQ at z $\approx$ 0.72, with RA = $179^\circ$.88 and Dec = $29^\circ$.24. In the $\gamma$-ray band, this source was first detected
by the Energetic Gamma Ray Experiment Telescope \citep{1995MNRAS.275..255T} and by VERITAS in the very high energy ($> 100$ GeV) \citep{2017ATel11075....1M}. The source exhibits optical variability and strong polarization \citep{10.1093/pasj/58.5.797}. During Fermi-LAT observations over 15 years, Ton 599 exhibited major flares in 2017 and January 2023. However, the brightest $\gamma$-ray flare was detected in January 2023, during which the SED was described by a BPL particle distribution for a one-zone leptonic model as reported by \citep{2024MNRAS.529.1356M}. In addition, different studies also reported the SED to be well described by a BPL distribution \citep{2020MNRAS.492...72P, 2024MNRAS.52711900R}. Moreover, in \cite{2024MNRAS.52711900R}, the authors have also fitted the SED with a power-law and log-parabola distribution of the particles. 

The paper is organized as follows: Section \ref{sec:obs} and \ref{sec:broadband} detail the data and broadband spectral analysis; Section \ref{sec:results} presents the results; and Section \ref{sec:summary} summarizes and discusses the findings. This work adopts the $\Lambda$CDM cosmological model with $H_0 = 71$ km s$^{-1}$ Mpc$^{-1}$, $\Omega_\Lambda = 0.73$, and $\Omega_M = 0.27$ \citep{Komatsu2011}.

\section{Observations and Data Analysis} \label{sec:obs}

\subsection{NuSTAR}

Nuclear Spectroscopic Telescope Array (NuSTAR) mission is the first space-based observatory in orbit to focus high energy X-rays in the band 3-79 keV. Launched by NASA on June 13, 2012, it extended the capability of imaging and spectroscopy above $\sim$ 10 keV. NuSTAR consists of two focal plane modules, FPMA and FPMB \citep{Harrison_2013}. The very high energy (VHE) detection of PKS 1441+25 by VERITAS from April 21 to April 28, 2015 triggered the X-ray observations \citep{2015ApJ...815L..22A}. NuSTAR observed the source on April 25, 2015 (MJD 57137) for a total exposure of 38.2 ks. The observations of PKS 1441+25 and Ton 599 used in this work are summarized in Table \ref{tab:obs}. 

We used `NUPIPELINE', which is integrated with ` HEASOFT-6.28' to generate the cleaned data files for further analysis. The package `XSELECT V2.4k' was used for plotting the event fits files \textit{(cl.evt)} in ds9. We have extracted a circular source region of 30 arcsec and a background region of 60 arcsec. The `NUPRODUCTS' task was used for the generation of spectrum files (PHA), ancillary files (ARF), and response matrix files (RMF). Finally, the spectra were grouped using the `GRPPHA' tool to ensure a minimum of 20 counts per bin for reliable spectral fitting.

\subsection{\textit{Swift}-XRT}

We analyzed \textit{Swift}-XRT observations that are nearly simultaneous with those from NuSTAR. The \textit{Swift}-XRT detector operates over an energy range of 0.2-10 keV featuring an effective area of 135 cm$^2$. Observations from both the instruments NuSTAR and \textit{Swift}-XRT were obtained from NASA's HEASARC archive\footnote{\url{https://heasarc.gsfc.nasa.gov/}}. We have used the software package `XRTPIPELINE'  which is integrated with `HEASOFT-6.28', for the generation of cleaned data and used  `XSELECT V2.4K' for plotting the event fits files \textit{(cl.evt)} in ds9. 

The data were taken in `PHOTON COUNTING' mode for both the sources. For PKS 1441+25, we extracted a source region with a radius of 20 pixels and a background region with a radius of 40 pixels. We checked for pile-up in the observations, and for the source Ton 599 pile-up was detected for which we extracted an annular region with an inner radius of 2 pixels and an outer region of 10 pixels, and a source contaminated free background region of 20 pixels. Light curves and spectra were extracted using `XSELECT v2.4k'. Ancillary response files (ARF) and response matrix files (RMF) were generated using the `{\it xrtmkarf'} and `{\it quzcif'} tools, respectively. Like NuSTAR, \textit{Swift}-XRT spectra were grouped such that there were at least 20 counts per bin.

\subsection{\textit{Swift}-UVOT}

The Ultra-Violet (UV) counterparts to the observations were obtained using \textit{Swift}-UVOT for both sources. It covers both the optical and UV bands of EM spectrum including three optical filters (B, V, U) and three UV filters (UVM1, UVM2, UVW2) \citep{2005SSRv..120...95R}. For PKS 1441+25, only the B, U, and UVW1 filter observations were available. We extracted source and background regions of 8 arcsec and 16 arcsec. For Ton 599, observations were available for all optical and UV filters, and we extracted source and background regions of radii 7 arcsec and 14 arcsec, respectively.

While the tool `UVOTSOURCE' was used for the magnitude calculation in the AB magnitude system and these were further corrected for interstellar reddening using the relation $\frac{A_v}{E(B-V)} = 3.1$, where A$_v$ is the Galactic extinction and $E(B-V)$ is the interstellar reddening. For PKS 1441+25, $E(B-V)$ = 0.038 mag, while for Ton 599, $E(B-V)$ = 0.019 mag \citep{schlafly2011measuring}. The fluxes were calculated from the magnitude values using photometric zero points and conversion factors obtained from \cite{2011AIPC.1358..373B} and \cite{10.1093/mnras/stw1516}. The flux values and corresponding energies from the UVOT observations were then converted into a PHA format compatible with `XSPEC version 12.11.1' \citep{1996ASPC..101...17A} using the tool `\textit{ftflx2xsp}'.

\subsection{\textit{Fermi}-LAT}
The source PKS 1441+25 showed significant flaring in the $\gamma$-ray band during January 2015, with flux doubling short time scale variability of $\sim$ 1.44 days \citep{2017MNRAS.470.2861S}. The strong emission of up to 250 GeV detected by VERITAS and MAGIC triggered follow-up observations with other instruments like NuSTAR, Swift and \textit{Fermi}-LAT. The large area telescope (LAT) operates in the energy range from 20 MeV to more than 300 GeV \citep{Atwood_2009}. To achieve near-simultaneous coverage with NuSTAR and Swift and ensure significant detection, we considered 4 months of \textit{Fermi}-LAT data from February 2024 to June 2024, with the NuSTAR observations occurring mid-interval.

For Ton 599, significant detection was obtained on June 26, 2021, enabling simultaneous observations with NuSTAR and Swift. For both sources, the analysis has been carried out with the help of open-source tools \textit{fermipy}\footnote{\url{https://fermipy.readthedocs.io/en/latest}}. The analysis was conducted with a region of interest (ROI) of 15$^{\circ}$, following the standard procedures outlined in \cite{2017ICRC...35..824W}. The energy range considered spanned from 100 MeV to 300 GeV, with evclass=128 and evtype=3. The latest instrument function (IRF), \lq\textit{P8R3 SOURCE V3}\rq, was utilized. For generating XML, the Galactic diffuse emission model \lq\textit{gll iem v07.fits}\rq and the isotropic background model \lq\textit{iso P8R3 SOURCE V3 v1.txt}\rq were employed. The XML model file, which includes all sources within the ROI, was created using the latest \textit{Fermi}-LAT 4FGL catalog \citep{2020ApJS..247...33A}. Additionally, the spectral models and parameters for the sources in the ROI were derived from the fourth \textit{Fermi} source catalog. Finally, the flux points and energies from the \textit{Fermi}-LAT observations were converted into a format readable by XSPEC (PHA format) using the tool \lq\textit{ftflx2xsp}\rq.

\subsection{\textit{VERITAS}}

For the very-high-energy (VHE) $\gamma$-ray analysis of PKS 1441+25, we used the TeV spectrum reported by \cite{2015ApJ...815L..22A}, with a total exposure time of 15 hrs. The flux points were corrected for extragalactic background light (EBL) absorption and converted to PHA format for XSPEC using \lq\textit{ftflx2xsp}\rq.  

We fitted the observations in XSPEC using a local broadband SED model, \textit{sscicon} \citep{10.1093/mnras/stae706}. For the spectral fit, we have used the \lq\textit{Tbabs}\rq model \citep{Wilms_2000} to account for the Galactic absorption. For PKS 1441+25, the hydrogen column density for the X-ray observations, ${N_\text{H}} = 3.14 \times 10^{20} $ $\text{cm}^{-2}$ was considered and kept fixed which was obtained in the LAB survey \citep{2005A&A...440..775K} while for Ton 599, the hydrogen column density for the X-ray observations, ${N_\text{H}} = 1.63 \times 10^{20}$ $\text{cm}^{-2}$ was considered and kept fixed\footnote{\url{https://heasarc.gsfc.nasa.gov/cgi-bin/Tools/w3nh/w3nh.pl}}. For the UV observations, the $N_\text{H}$ value was fixed at 0 as the fluxes were de-redenned initially. The errors are estimated at a 90\% confidence level with the standard XSPEC method of $\chi^2$ error calculation\footnote{\url{https://heasarc.gsfc.nasa.gov/xanadu/xspec/manual/XSerror.html}}.

\begin{table*}[ht]
\renewcommand{\arraystretch}{1.5} 
\caption {\label{tab:obs} Summary of \textit{NuSTAR} and \textit{\textit{Swift}}-XRT/UVOT observations.}
\setlength{\tabcolsep}{2pt}

\centering
\begin{tabularx}{\textwidth}{ccccccccccc}
\hline
& & \textit{NuSTAR}& & & &\textit{Swift}-XRT/UVOT& & & \\ 
\hline
Source Name &  Obs ID   & Date and Time  & MJD & Exposure (ks) &  Obs ID   & Date and Time  & Exposure (XRT/UVOT) (ks)\\
\hline
 
 PKS 1441+25 &  90101004002 & 2015-04-25 T02:41:07 & 57137 & 38.24 &   
 
        00040618014 & 2015-04-25 T09:13:07 & 1.98/1.98 \\
 
 Ton 599 & 60463037004  & 2021-06-25 T12:46:09 & 59390 & 17.59 &

         00036381062 & 2021-06-25 T04:55:36 & 0.8/0.78 \\
          
\hline

\end{tabularx}
\end{table*}

\section{Broadband Spectral Analysis}\label{sec:broadband}
Our Broadband SED model assumes a leptonic distribution with a single emitting region characterized by several physical parameters, along with an isotropic electron distribution n($\gamma$) \citep{10.1093/mnras/stae706, hota2024multiwavelengthstudyextremehighenergy, tantry2024probingbroadbandspectralenergy}. The model does not incorporate synchrotron self-absorption and hence the actual flux at low (i.e. radio) frequencies will be significantly lower. As the sources are FSRQs, to ensure contributions from external photon fields, we also considered other parameters such as $BBfrac$ (fraction of photon contribution from the black body)  and $BBtemp$ (temperature of the emission region), depending on the location. The contribution of seed photons to the EC process could be from the accretion disk, BLR, or a dusty torus \citep{2002ApJ...575..667D,10.1111/j.1365-2966.2009.15397.x, Dermer_2009, 2024MNRAS.52711900R}. To fit the broadband spectra we have adopted various electron distribution models that can convolve with the model \citep{Hota_2021, Khatoon_2022, 10.1093/mnras/stae706, hota2024multiwavelengthstudyextremehighenergy, tantry2024probingbroadbandspectralenergy}. In particular, we have used the synchrotron flux relation mentioned in \cite{Hota_2021}, \cite{Khatoon_2022}, \cite{10.1093/mnras/stae706} and \cite{hota2024multiwavelengthstudyextremehighenergy} given by,   

\begin{equation}\label{flux}
    F_{\text{syn}} (\epsilon) = \frac{\delta^3 (1+z)}{d_L^2} V \mathbb{A} \int_{\xi_\text{min}}^{\xi_\text{max}} f \left( \frac{\epsilon}{\xi^2} \right) n(\xi) d\xi
\end{equation}

\noindent where $d_L$ is the luminosity distance and V is the volume of the emission region, $ \mathbb {A} = \frac{\sqrt{3} \pi e^3 B}{16 m_e c^2\sqrt\mathbb{{C}}}$ and $f(x)$ is the synchrotron emissivity function. Instead of using the electron's Lorentz factor $\gamma$, the electron energy distribution is represented by $n (\xi)$. The transformation is given by $\xi=\gamma \sqrt \mathbb{C}$, where $ \mathbb {C}=1.36 \times 10^{-11}\frac{\delta B}{1+z}$ keV with $\delta$ as the doppler factor, B as the magnetic field, and z being the redshift of the source. Note that $\xi$ represents $\gamma$ and $\xi^2$ has dimension of keV. 

For the fitting of $\gamma$-ray observations as well as some in X-rays, the EC and SSC mechanisms have been employed whose fluxes have been determined as a function of $F_\text{syn} (\epsilon)$ and $n(\xi)$. The detailed flux expressions have already been given in \cite{Sahayanathan_2018}. The \textit{sscicon} consists of parameters like bulk Lorentz factor $\Gamma$, redshift $z$, magnetic field $B$, viewing angle $i$, emission region size $R$, $BBtemp$, $BBfrac$, jet power $P_j$ and other parameters from the user given particle distribution  $n(\xi)$ including normalization $K$ \citep{10.1093/mnras/stae706, hota2024multiwavelengthstudyextremehighenergy, tantry2024probingbroadbandspectralenergy}. Mathematically, total jet power is given as \citep{Ghisellini_2014},

\begin{equation*}
    P_j = 2 \pi R^2 \Gamma B U'_j    
\end{equation*}
where, factor 2 accounts for the two sided jet, $R$ is the emission region size, $U'_j$ is the energy density of electron ($j = e$), protons ($j = p$), magnetic field ($j = B$), and radiation ($j = \text{rad}$) in the co-moving frame.

Below, we briefly discuss the particle distributions used for this study.  

\subsection{{Logparabola model}}
The particle density for a log-parabolic model is given by,

\begin{equation}\label{lp} 
    n({\xi})=K \left (\frac{\xi}{\xi_r} \right)^{- \alpha - \beta \text{log} \left(\frac{\xi}{\xi_r} \right)} 
\end{equation}

\noindent In this expression, $\alpha$ represents the particle spectral index at the reference energy $\xi_r$, while $\beta$ and $K$ denote the spectral curvature parameter and normalization, respectively. During the spectral fit, $\xi_r^2$ was fixed at 1 keV, while $\alpha$, $\beta$, and the normalization $K$ were treated as free parameters.

\subsection{Broken Power Law}
We applied a Broken Power Law (BPL) distribution to model the particle spectrum,

\begin{equation}\label{bpl}
n(\xi)=
\begin{cases}
    K (\xi/1\sqrt{{\text{keV}}})^{-p} & \text{for} \hspace{3pt} \xi < \xi_\text{break} \\
    K \xi^{q-p}_\text{break}(\xi/1\sqrt{{\text{keV}}})^{-q} & \text{for} \hspace{3pt} \xi > \xi_\text{break}  
\end{cases}
\end{equation}

\noindent Here, $\xi_\text{break}$ represents the break energy, $p$ is the electron spectral index for $\xi < \xi_\text{break}$, and $q$ is the electron spectral index for $\xi > \xi_\text{break}$. The transformation is defined by $\xi = \gamma \sqrt{\mathbb{C}}$. This broken power-law particle distribution is considered valid only within the range $\xi_\text{min} < \xi < \xi_\text{max}$, where $\xi_\text{min} = \gamma_\text{min} \sqrt{\mathbb{C}}$ and $\xi_\text{max} = \gamma_\text{max} \sqrt{\mathbb{C}}$.

\subsection{Energy dependent time-scale models:}

We employed models that incorporate energy-dependent escape or acceleration timescales to help explain the curvature observed in the spectrum \citep{Hota_2021,Khatoon_2022,10.1093/mnras/stae706, hota2024multiwavelengthstudyextremehighenergy, tantry2024probingbroadbandspectralenergy}.

\subsubsection{Energy Dependent Diffusion Model (EDD)}

In this model, diffusion occurs in a region with a tangled magnetic field, causing the escape timescale to depend on the electron's gyration radius. This makes the escape timescale ($\tau_\text{esc}$) energy-dependent, given by

\begin{equation} \tau_\text{esc} = \tau_{\text{esc},R} \left(\frac{\gamma}{\gamma_R} \right)^{-k} \end{equation}

\noindent where $\tau_{\text{esc},R}$ is the escape timescale when the electron energy is $\gamma_{R}mc^2$, and $k$ is the index describing the power-law dependence on energy.

The escape timescale $\tau_\text{esc}$ cannot exceed the free-streaming limit, so this relationship is valid only for $\gamma < \gamma_R$, where $\gamma_R$ is the energy at which this limit is reached.

Assuming $\gamma_R$ is much larger than any $\gamma$ of interest and neglecting synchrotron losses, the resulting electron energy distribution is,

\begin{equation} n(\xi) = Q_0 \tau_\text{acc} \sqrt{\mathbb{C}} \xi^{-1} \exp \left[ -\frac{n_R}{k} \left( \left(\frac{\xi}{\xi_R} \right)^k - \left( \frac{\xi_0}{\xi_R} \right)^k \right) \right] \end{equation}

\noindent where $\xi_R = \sqrt{\mathbb{C}} \gamma_R$, $\xi_0 = \sqrt{\mathbb{C}} \gamma_0$, and $\eta \equiv \frac{\tau_\text{acc}}{\tau_{\text{esc},R}}$.

The distribution can be conveniently rewritten as,

\begin{equation} n(\xi) = K \xi^{-1} \exp \left[-\frac{\psi}{k} \xi^{k} \right] \end{equation}

\noindent where $K$, $\psi$, and $k$ are free parameters. It can be shown that,

$$\psi = \eta_R (\mathbb{C}\gamma^{2}_{R})^{-k/2}=\eta_R \xi_{R}^{-k}$$

and that the normalization is given by,
    
$$K = Q_0 \tau_{acc} \text {exp} \left [\frac{n_R}{k} \left(\frac{\xi_0}{\xi_{R}} \right)^k \right]$$

\subsubsection{Energy dependent acceleration (EDA) model}
We consider the possibility that the acceleration timescale is energy-dependent, assuming it follows the functional form,

\begin{equation} \tau_\text{acc} = \tau_{\text{acc},R} \left( \frac{\gamma}{\gamma_R} \right)^k \end{equation}

\noindent where $\tau_{\text{acc},R}$ is the acceleration timescale when the electron energy is $\gamma_R mc^2$, and $k$ is the index describing the energy dependence. As in the EDD model, the resulting electron energy distribution is given by,

\begin{equation} \small n(\xi) = Q_0 \tau_\text{acc} \sqrt{\mathbb{C}} \xi_R^{-k} \xi^{k-1} \exp \left[ -\frac{n_R}{k} \left( \left( \frac{\xi}{\xi_R} \right)^k - \left( \frac{\xi_0}{\xi_R} \right)^k \right) \right] \end{equation}

\noindent which can be recast as,

\begin{equation} n(\xi) = K \xi^{k-1} \exp \left[ -\frac{\psi}{k} \xi^k \right] \end{equation}

\noindent where,\\
$$\psi = \eta_R (\mathbb{C}\gamma^{2}_{R})^{(-k/2)} = \eta_R \xi_{R}^{-k}$$

\noindent K is the normalization which is given by,
    
$$K = Q_0 \tau_{acc} \xi_{R}^{-k} \text{exp} \left [\frac{n_R}{k} \left(\frac{\xi_0}{\xi_{R}} \right )^k \right]$$

\section{Results} \label{sec:results}

\begin{figure*}
\centering
    \includegraphics[width=.34\textwidth, angle = 270]{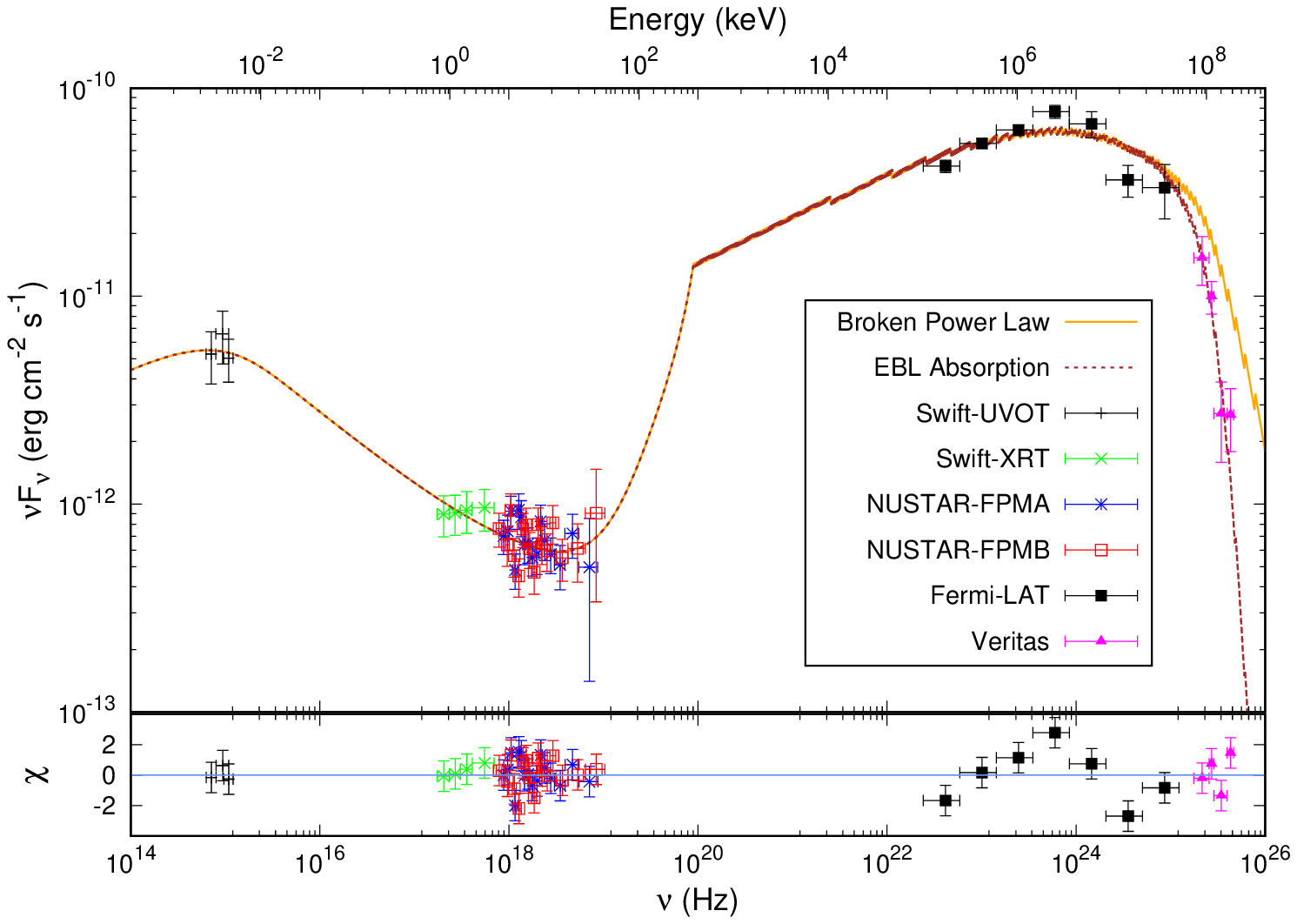}\hfill
    \includegraphics[width=.34\textwidth, angle = 270]{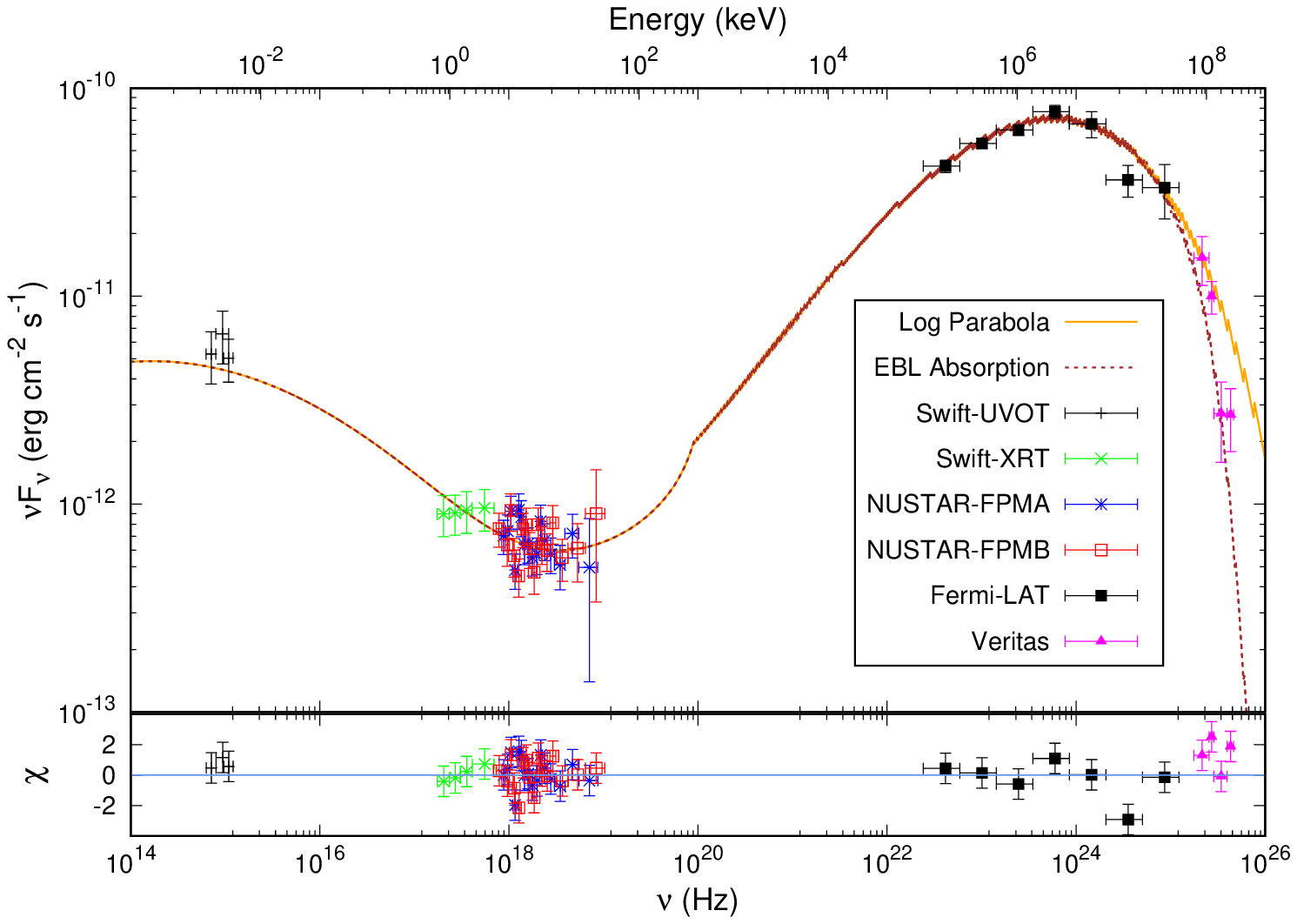}\hfil
    \includegraphics[width=.34\textwidth, angle = 270]{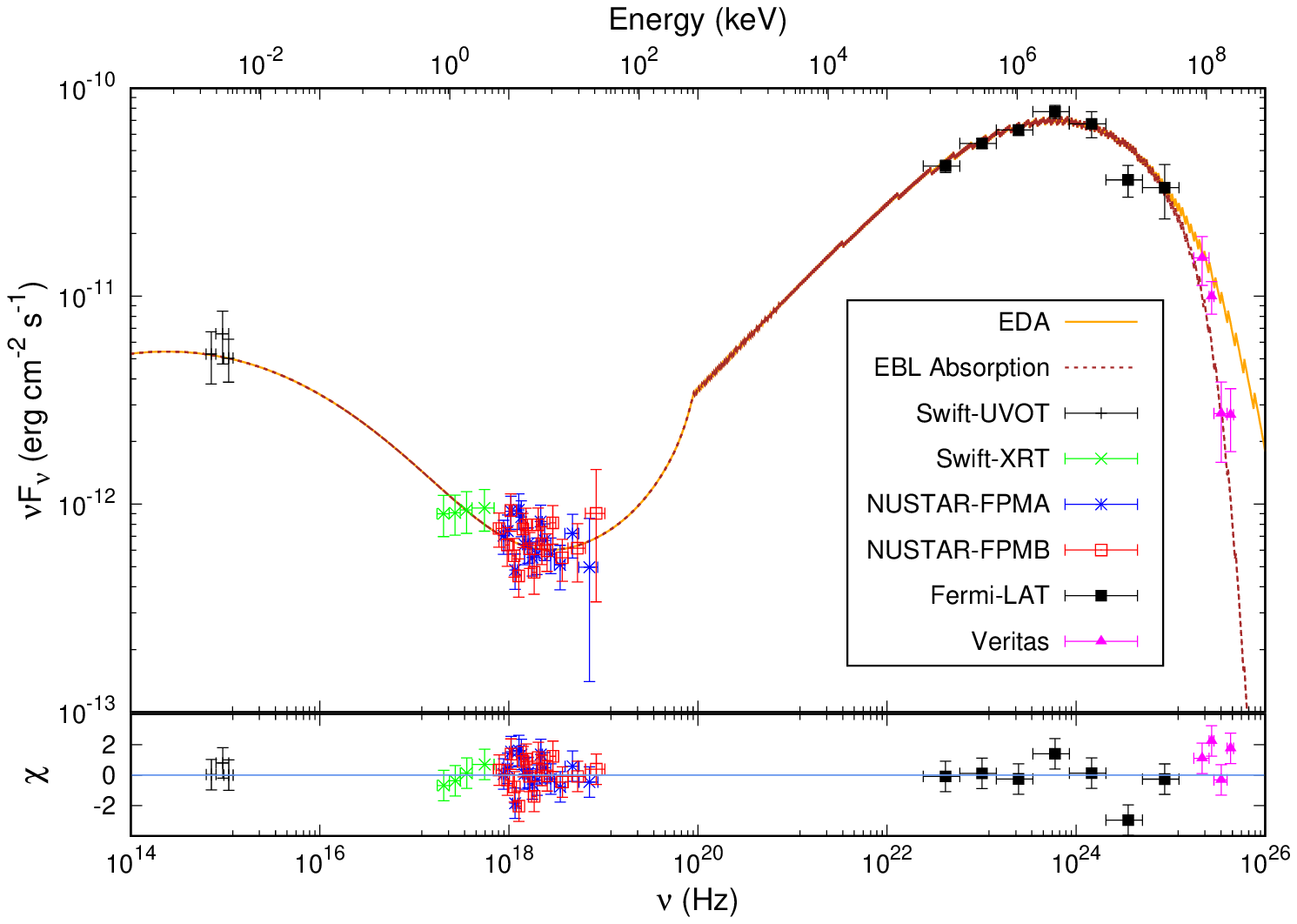}\hfill
    \includegraphics[width=.34\textwidth, angle = 270]{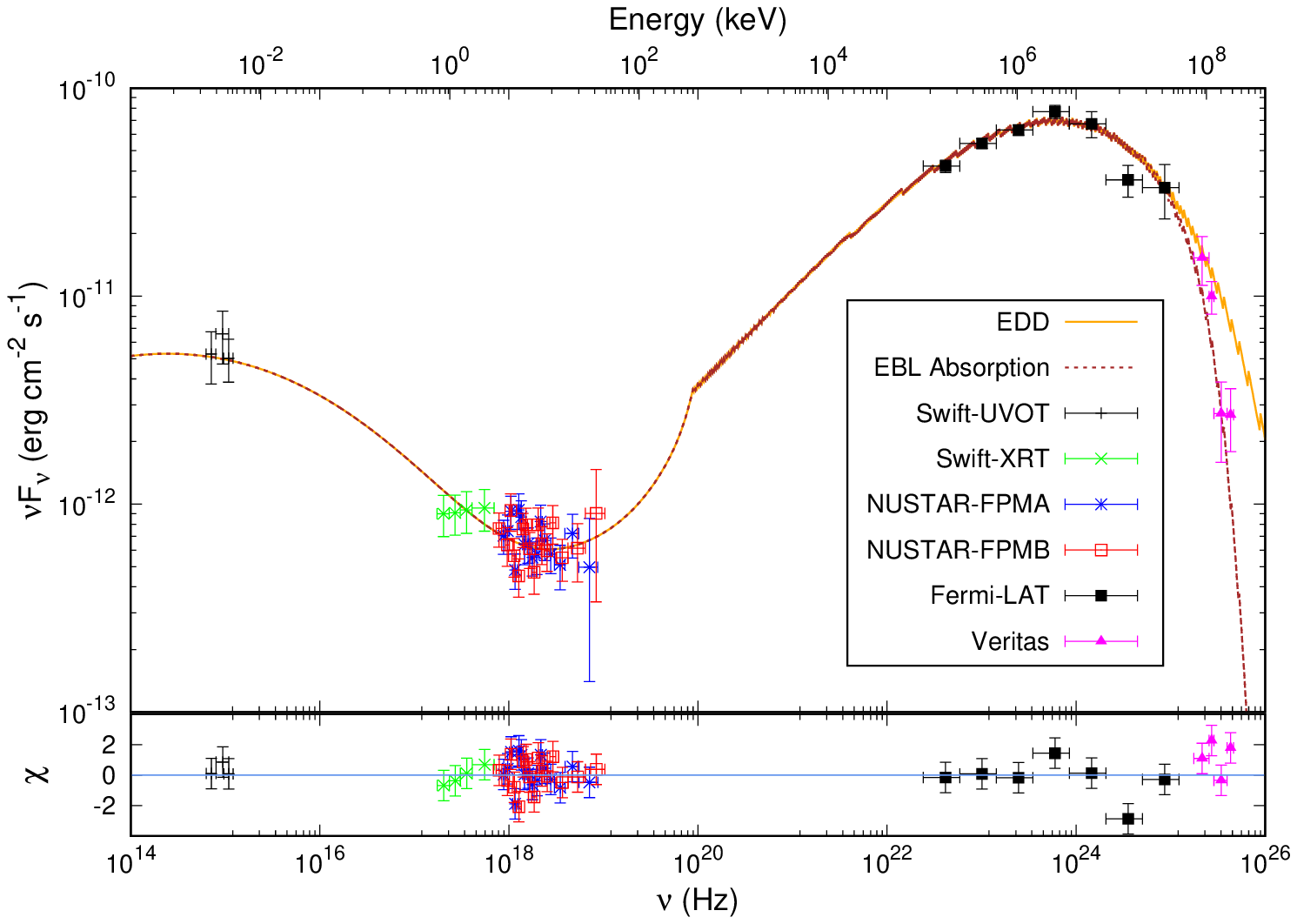}
    \caption{Broadband SED plot of PKS 1441+25 for a $\gamma_\text{min} = 10$, $\Gamma= 20$, $R$ = 10$^{17}$ cm, BBtemp = 10$^{4}$ K for Broken Power law (Top left) and Log Parabola (Top right), EDA (Bottom left) and EDD (Bottom right).}
    \label{sed_pks}
\end{figure*}

\begin{figure*}
\centering
    \includegraphics[width=.34\textwidth, angle = 270]{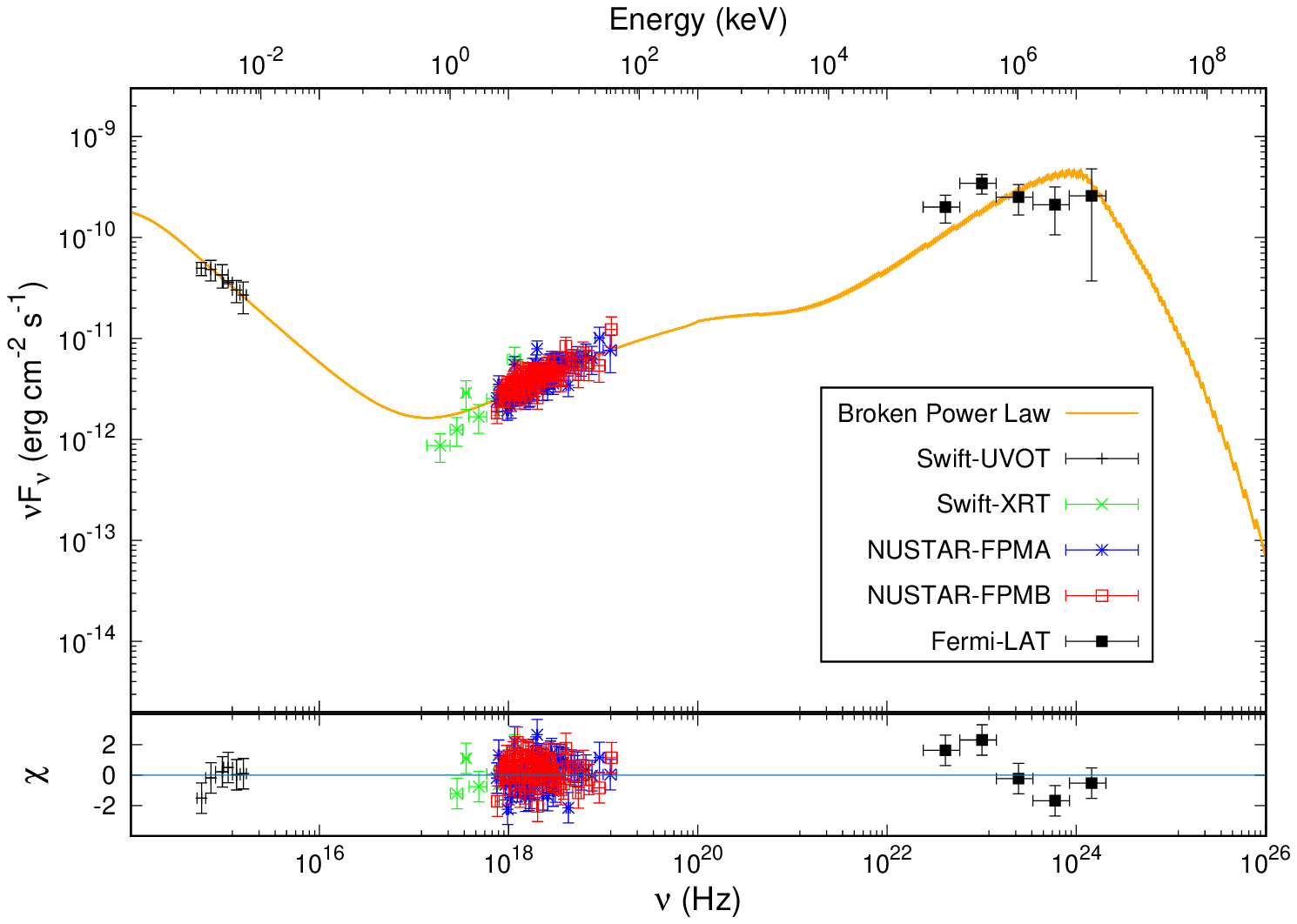}\hfill
    \includegraphics[width=.34\textwidth, angle = 270]{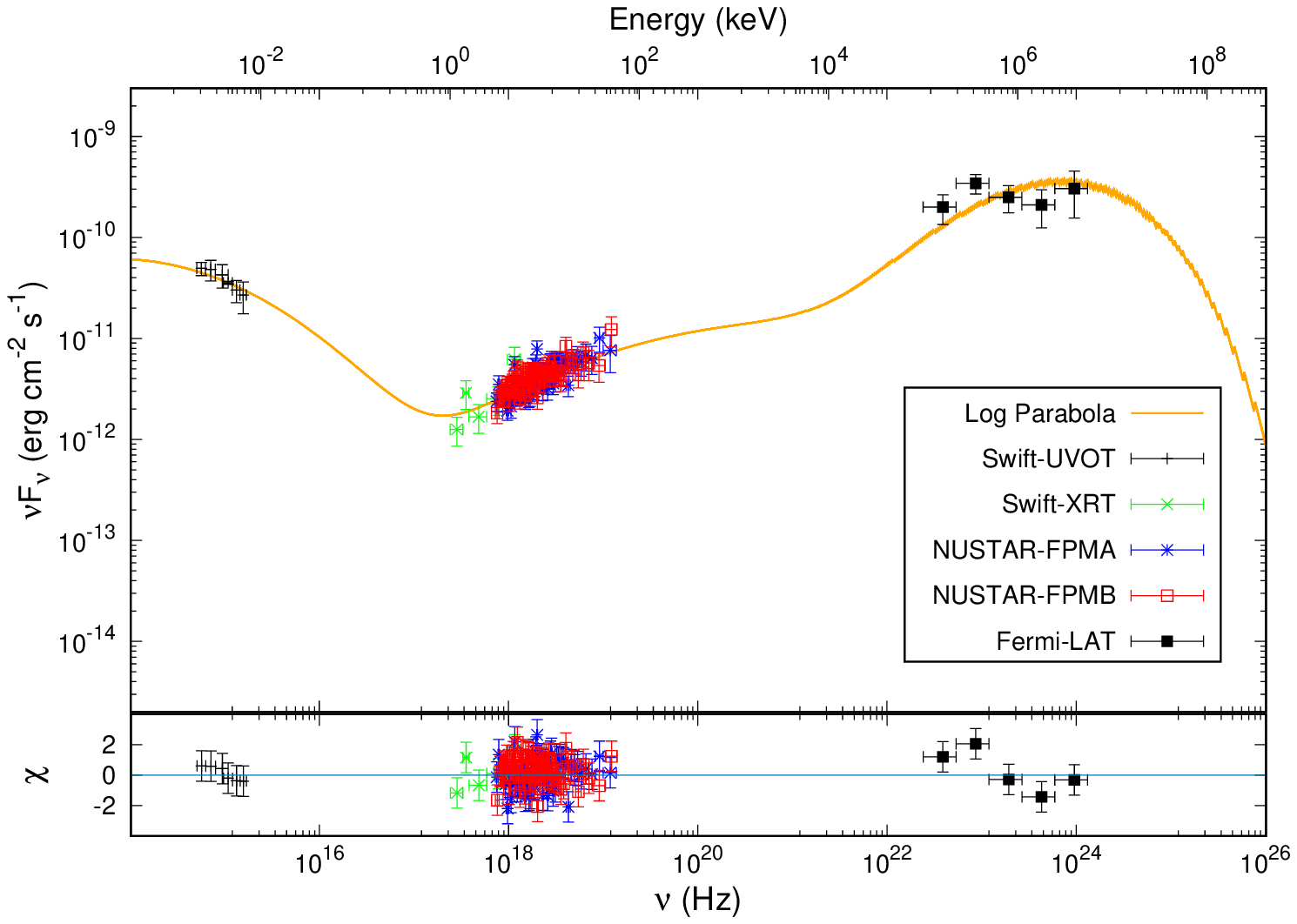}\hfill
    \includegraphics[width=.34\textwidth, angle = 270]{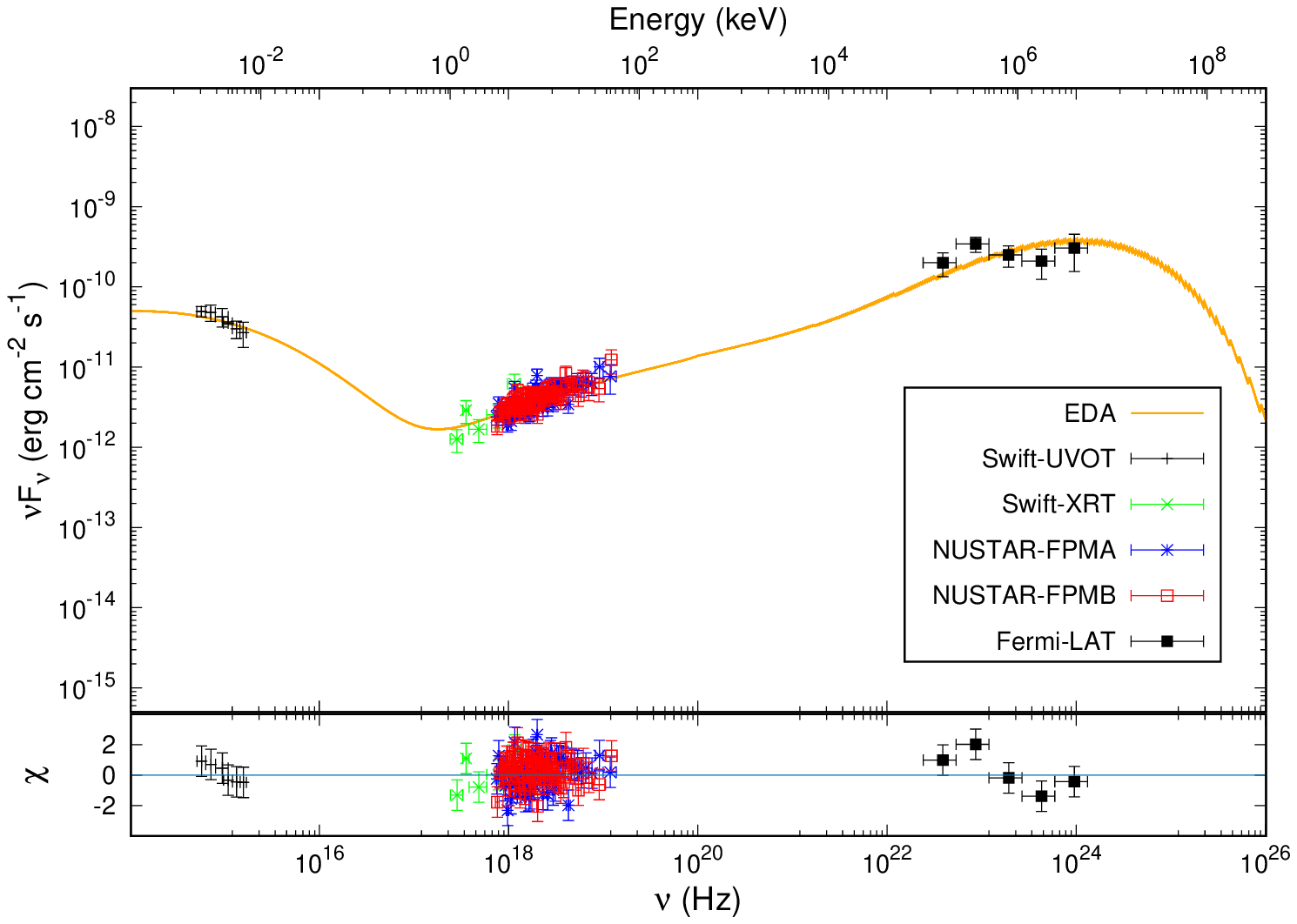}\hfill
    \includegraphics[width=.34\textwidth, angle = 270]{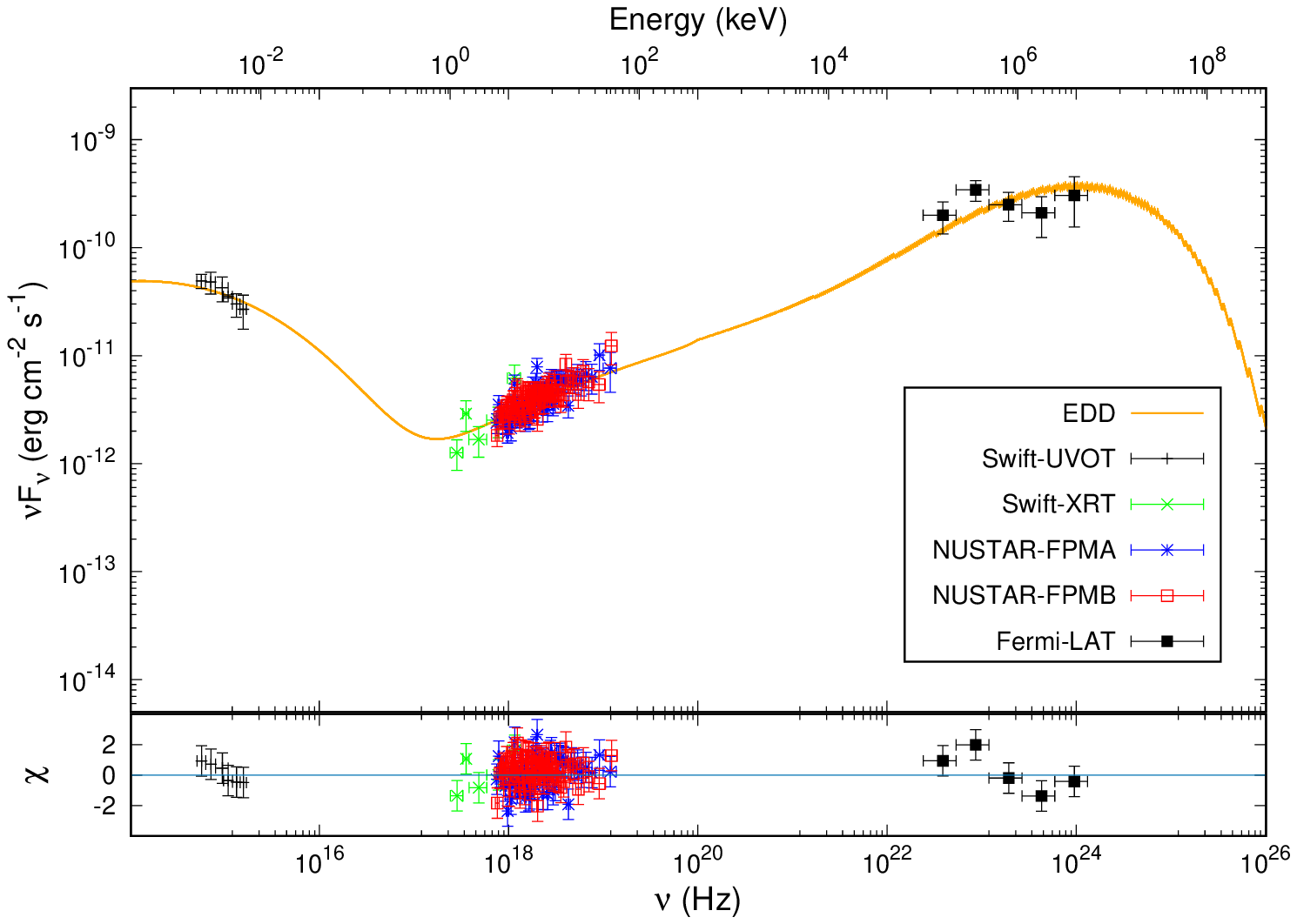}
    \caption{Broadband SED plot of Ton 599 for a $\gamma_\text{min} = 10$, $\Gamma = 20$, $R$ = 10$^{17}$ cm, BBtemp = 10$^{4}$ K for Broken Power law (Top left) and Log Parabola (Top right), EDA (Bottom left) and EDD (Bottom right).}
    \label{sed_ton}
\end{figure*}

We have analyzed and fitted the multi-wavelength observations of two FSRQ sources PKS 1441+25 and Ton 599 which are listed in Table \ref{tab:obs}. For PKS 1441+25, the considered observations were in the historically high flux state \citep{2015ApJ...815L..22A}. We fitted the observations with Log parabola, broken power law, EDA, and EDD particle distributions. During the fit, we fixed the region size at $ R = 10^{17}$ cm and viewing angle $i=0^\circ$ and allowed the other parameters to vary. In Figure \ref{sed_pks} and \ref{sed_ton}, we have shown the broadband SED fitting with all the models for a Lorentz factor, $\Gamma = 20$, $\gamma_\text{min} = 10$, $R$ = 10$^{17}$ cm, $BBtemp$ = 10$^{4}$ K, while the best-fit parameters are reported in Table \ref{tab:bestfit}. The underlying radiative processes considered for the fitting are the synchrotron and EC. 

\begin{table*}[h!]
\renewcommand{\arraystretch}{1.5} 
\caption {\label{tab:bestfit} The best-fit parameters obtained for all the particle distributions considered with the one-zone leptonic model at $\Gamma = 20$, $ \gamma_\text{min} = 10$, size of the region at, $R=10^{17}$ cm, and Temperature at, $T = 10^4$ K. Instead of the magnetic field, the equipartition parameter $\sigma$ has been used. The minimum electron energy and the external field contribution are given by $\gamma_\text{min}$ and $BBfrac$ respectively. The jet power $P_j$ in logarithmic scale with the unit of erg s$^{-1}$.}    
\setlength{\tabcolsep}{8pt}
\begin{tabularx}{\textwidth}{cccccccccccc}

\hline
& & & & & Log Parabola & & &   \\
\hline

& Source & $\alpha$  & $\beta$ & $\sigma = U_B/U_e$ & BBfrac ($10^{-8}$) & log $P_{j}$  & $\chi^{2}_\text{red}(\text{dof})$ \\
\hline
 
& PKS 1441+25 & 3.9$^{+0.2}_{-0.1}$  & 0.31$^{+0.02}_{-0.03}$ & 0.04 & 6.3$^{+3.1}_{-1.4}$ & 46.54$^{+0.01}_{-0.01}$ & 1.17 ({45})\\ 

& Ton 599 & 5.9$^{+0.3}_{-0.4}$  & 0.83$^{+0.09}_{-0.11}$ & 0.49 & 11.4$^{+2.7}_{-2.7}$ & 45.898$^{+0.006}_{-0.006}$ & 0.87 ({136})\\ 
  
\hline

& & & & &  Broken Power Law  & & &   \\
\hline
& Source  & $p$ & $q$ & $\xi^2_\text{break}$ ${\text{(eV)}}$ & $\sigma = U_B/U_e$ & BBfrac ($10^{-8}$)& log $P_{j}$ & $\chi^{2}_\text{red}(\text {dof})$ &   \\
\hline

& PKS 1441+25 & 2.55$^{+0.04}_{-0.05}$  & 3.73$^{+0.14}_{-0.13}$ &  3.3$^{+0.1}_{-0.1}$ & 0.04 & 10.16$^{+8.05}_{-4.40}$ & 47.18$^{+0.01}_{-0.01}$ & 1.24 (44)  \\

& Ton 599 & 1.7$^{+0.2}_{-0.2}$ & 4.6$^{+0.4}_{-0.3}$ & 0.266$^{+0.002}_{-0.002}$ & 0.9 & 14.9$^{+4.8}_{-4.3}$ &  46.235$^{+0.005}_{-0.005}$ & 0.93 (135)  \\      
 
\hline

& & & & & EDA  & & &   \\

\hline
& Source & $\psi$& k & $\sigma = U_B/U_e$ & BBfrac ($10^{-8}$) & log $P_{j}$ & $\chi^{2}_\text{red}(\text {dof})$\\
\hline
 
 & PKS 1441+25 & 3.3$^{+0.2}_{-0.2}$  & 0.13$^{+0.01}_{-0.01}$ & 0.03 & 5.9$^{+1.8}_{-1.6}$ & 46.80$^{+0.01}_{-0.01}$ &  1.11 (45)  \\  

 & Ton 599 & 6.40$^{+1.33}_{-0.91}$  & 0.27$^{+0.06}_{-0.03}$ & 0.14 & 9.5$^{+2.9}_{-2.3}$ & 46.404$^{+0.006}_{-0.007}$ &  0.89 (136)  \\        
\hline

& & & & & EDD  & & &  \\

\hline
& Source &$\psi$ & k & $\sigma = U_B/U_e$ & BBfrac ($10^{-8}$) &  log $P_{j}$ & $\chi^{2}_\text{red}(\text {dof})$\\
\hline
 
&PKS 1441+25 & 3.2$^{+0.2}_{-0.1}$  & 0.13$^{+0.01}_{-0.01}$ & 0.03 & 5.6$^{+1.8}_{-1.5}$ & 46.83$^{+0.01}_{-0.01}$ &  1.11 (45)  \\  

& Ton 599 & 6.3$^{+0.1}_{-0.4}$ & 0.30$^{+0.07}_{-0.04}$ & 0.14 & 9.5$^{+2.2}_{-2.7}$ & 46.483$^{+0.006}_{-0.007}$ &  0.89 (136)  \\       
\hline
\end{tabularx}

\end{table*}

For the source PKS 1441+25 in Figure \ref{sed_pks}, the UV and X-ray observations were fitted with the synchrotron component, while the $\gamma$-ray observations are described by the EC component. The contribution from the BLR is considered to be in the form of a black body, for which the temperature considered is $T\sim 10^4$ K \citep{2006LNP...693...77P}. The fraction of contribution from the black body is given by $BBfrac$ was kept free, while $\gamma_\text{min}$ was kept frozen at 10. 

\begin{figure*}
\centering
    \includegraphics[width=.5\textwidth]{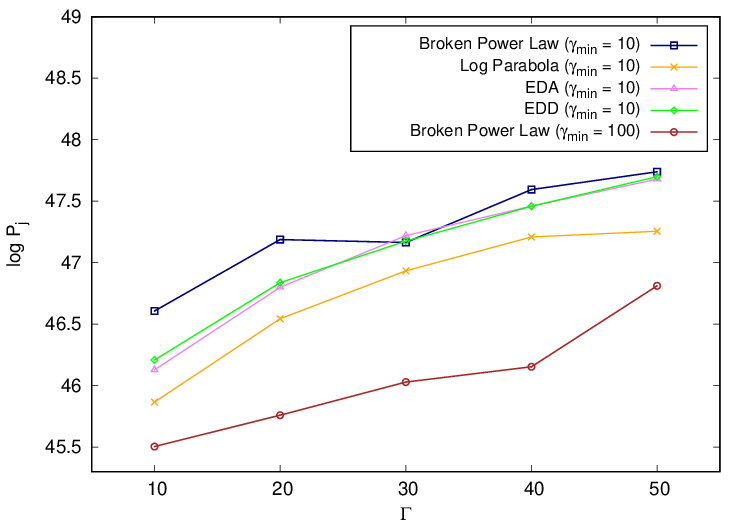}\hfill
    \includegraphics[width=.5\textwidth]{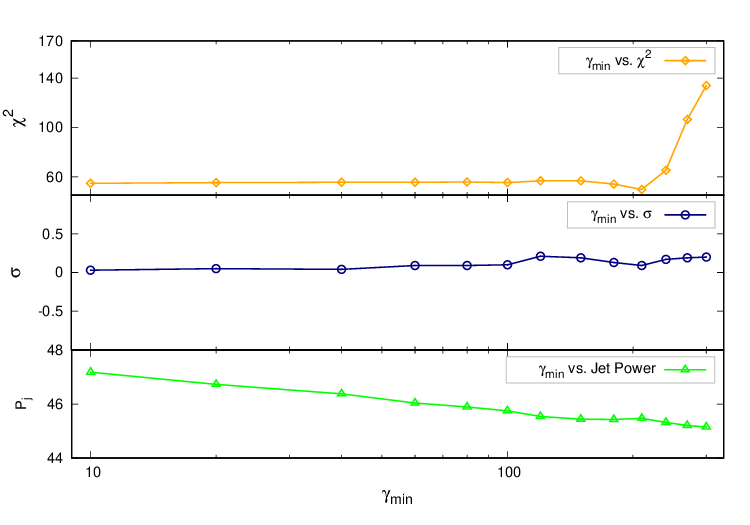}
    \caption{Variation of jet power with $\Gamma$ at a $T=10^4 K$, $R = 10^{17}$ cm (left), and variation of $\chi^2$ (top panel), equipartition value $\sigma$ (middle panel) and jet power $P_j$ (lower panel) for different values of $\gamma_\text{min}$ (right) for BPL distribution for PKS 1441+25.}
    \label{lf_vs_jp}
\end{figure*}

Based on previous studies, we first adopt $\Gamma=20$ as a representative value and subsequently explore a broader range of Lorentz factors reported in the literature to examine their impact on the inferred jet powers. Accordingly, following earlier works on PKS 1441+25 and FSRQs in general  \citep{Ahnen_2015, 2015ApJ...815L..22A, 10.1093/mnras/stv055}, we fitted the SED for different values of $\Gamma$, as shown in the left panel of Figure \ref{lf_vs_jp}, and subsequently estimated the jet power for all particle distributions considered. Furthermore, we explored the range of $\gamma_\text{min}$  values that yield statistically acceptable fits, as shown in the right panel of Figure \ref{lf_vs_jp}.

\begin{figure*}[ht!]
\centering
    \includegraphics[width=.5\textwidth]{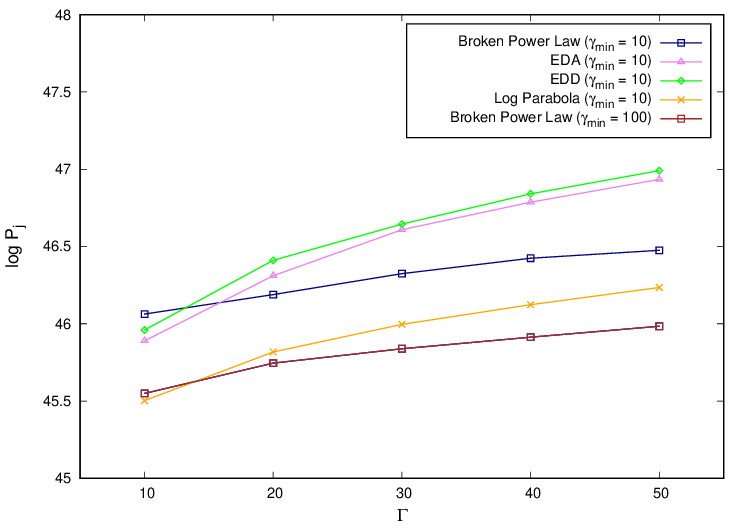}\hfill
    \includegraphics[width=.5\textwidth]{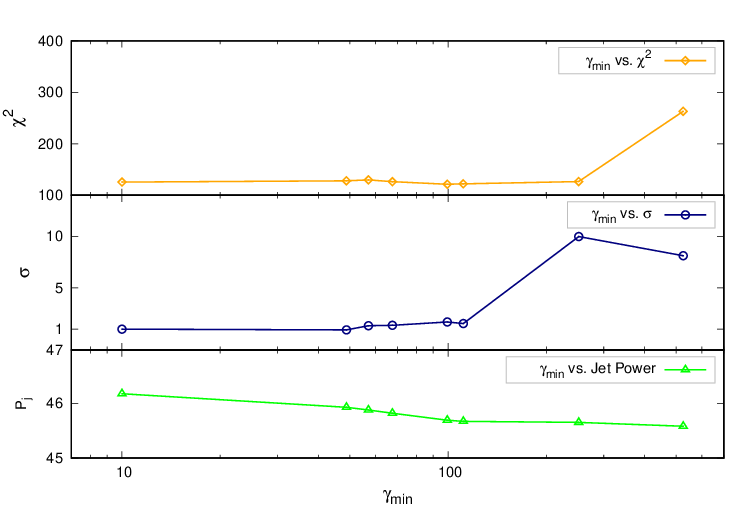}
    \caption{Variation of jet power with $\Gamma$ at a $T=10^4 K$, $R = 10^{17}$ cm (left), and variation of $\chi^2$ (top panel), equipartition "$\sigma$" (middle panel) and jet power $P_j$ (lower panel) for different values of $\gamma_\text{min}$ (right) for BPL distribution for Ton 599.}
    \label{lf_vs_jp_ton}
\end{figure*}

For Ton 599, the UV observations are described by the synchrotron component, while the X-ray emission can be modeled with SSC and EC.  However, fitting the X-rays with EC required $\gamma_\text{min} < 1$, which is physically implausible since $\gamma_\text{min}$ denotes the minimum Lorentz factor of relativistic electrons. We examined the variation of jet power for different Lorentz factors ($\Gamma$) for all the particle energy distributions considered (Figure \ref{lf_vs_jp_ton}, left panel)\citep{2024MNRAS.529.1356M, 2024MNRAS.52711900R}. In addition to the total jet power calculation, we calculated the power carried by each component as reported in Tables \ref{gammamin=10_pks} and \ref{gammamin=10_ton}. 

\begin{table}[hb!]
\renewcommand{\arraystretch}{1.5} 
\caption {Jet Power components in erg s$^{-1}$ for all the particle distribution for PKS 1441+25 for $\gamma_\text{min} = 10$, $\Gamma = 20$, $R=10^{17}$ cm.}  
    \centering
    \begin{tabular}{cccccc}
        \hline
         Model &  $P_e$ & $P_p$ & $P_B$ & $P_T$ \\
         \hline
         BPL & $2.3 \times 10^{45}$ & $1.5 \times 10^{47}$ & $8.5 \times 10^{43}$ & $1.5 \times 10^{47}$ \\
         LP & $1.34 \times 10^{45}$ & $3.35 \times 10^{46}$  & $6.6 \times 10^{43}$ & $3.49 \times 10^{46}$ \\
         EDA & $1.77 \times 10^{45}$ & $6.22 \times 10^{46}$ & $6.6 \times 10^{43}$ & $6.4 \times 10^{46}$ \\
         EDD & $1.84 \times 10^{45}$ & $6.69 \times 10^{46}$ & $6.3 \times 10^{44}$ & $6.88 \times 10^{46}$ \\
        \hline
    \end{tabular}
    \label{gammamin=10_pks}
\end{table}

\begin{table}[hb!]
\renewcommand{\arraystretch}{1.5} 
\caption {Jet Power components in erg s$^{-1}$ for all the particle distribution for Ton 599 for $\gamma_\text{min} = 10$, $\Gamma = 20$, $R=10^{17}$ cm.}  
    \centering
    \begin{tabular}{cccccc}
        \hline
         Model &  $P_e$ & $P_p$ & $P_B$ & $P_T$ \\
         \hline
         BPL & $5.6 \times 10^{44}$ & $1.3 \times 10^{46}$ & $2.3 \times 10^{45}$ & $1.6 \times 10^{46}$ \\
         LP & $8.5 \times 10^{44}$ & $5.4 \times 10^{45}$  & $4.2 \times 10^{44}$ & $6.6 \times 10^{45}$ \\
         EDA & $1.5 \times 10^{45}$ & $1.8 \times 10^{46}$ & $1.2 \times 10^{44}$ & $2.04 \times 10^{46}$ \\
         EDD & $1.4 \times 10^{45}$ & $2.4 \times 10^{46}$ & $1.9 \times 10^{44}$ & $2.5 \times 10^{46}$ \\
        \hline
    \end{tabular}
    \label{gammamin=10_ton}
\end{table}

\section{Summary and Conclusions} \label{sec:summary}
We collated broadband data of PKS 1441+25 and Ton 599 during April 2015 and June 2021 respectively using \textit{Swift}-XRT/UVOT, NuSTAR, \textit{Fermi}-LAT and VERITAS, and modeled their spectra with a single-zone leptonic framework that incorporates synchrotron, synchrotron self-Compton (SSC), and external Compton (EC) processes. The fits were performed using four non-thermal electron energy distributions: broken power law (BPL), log-parabola (LP), energy-dependent diffusion timescale (EDD), and energy-dependent acceleration timescale (EDA). All models provided statistically acceptable fits (Table \ref{tab:bestfit}). 

We estimated and reported the jet powers of both FSRQs for all the models considered as shown in the left panel of Figure \ref{lf_vs_jp} and \ref{lf_vs_jp_ton}. Under each model (left panels, Figure \ref{lf_vs_jp} and \ref{lf_vs_jp_ton}), for Lorentz factors ($\Gamma$) between 10 and 50, the jet powers ranged from $10^{45} - 10^{48}$ erg s$^{-1}$ with the highest powers corresponding to the largest $\Gamma$. 

To ensure physically acceptable solutions, we analyzed the variation of $\chi^2$ with $\gamma_\text{min}$ (Figure \ref{lf_vs_jp} and \ref{lf_vs_jp_ton}, right panels). Although the X-ray spectra could be fit with either SSC or EC components, EC fits yielded unphysical $\gamma_\text{min}$ and equipartition values, so we retained only SSC fits for the X-ray band. Unlike the results of \cite{10.1093/mnras/stae706}, the jet powers in our analysis did not show a strong dependence on $\gamma_\text{min}$. Among the electron distributions, LP, EDA, and EDD models produced systematically higher jet powers which is nearly same as BPL model, reflecting the effect of intrinsic curvature in the particle spectrum.

For PKS 1441+25, the intrinsic curvature models for a $\Gamma = 20$ and a $\gamma_\text{min} \sim 10$, show powers higher than $\sim 10^{46}$ erg s$^{-1}$. For BPL distribution, the jet power was high as $10^{47}$ for a $\gamma_\text{min} \sim 10$, which reduced to $ \sim 10^{45}$ (varies by an order of $\sim$ 2) for a $\gamma_\text{min} \sim 100$ for an acceptable equipartition value. However, this doesn't seem to be the same in the case of FSRQ source Ton 599. Here, the jet powers of all the models for a $\Gamma = 20$, and a $\gamma_\text{min} \sim 10$ show power in the range $\sim 10^{45.5} - 10^{46.5}$ erg s$^{-1}$ and for the BPL distribution the jet powers vary by an order of $<$ 1, when $\gamma_\text{min}$ varies from 10 to 100. 

\begin{figure*}
\centering
    \includegraphics[width=.5\textwidth]{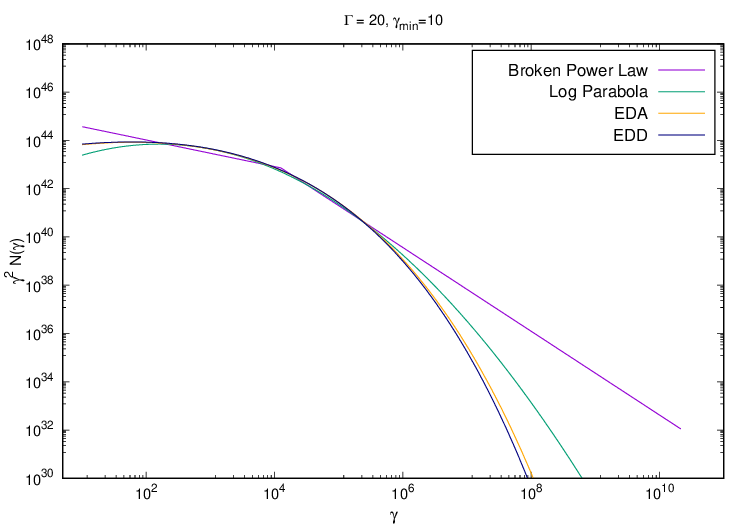}\hfill
    \includegraphics[width=.5\textwidth]{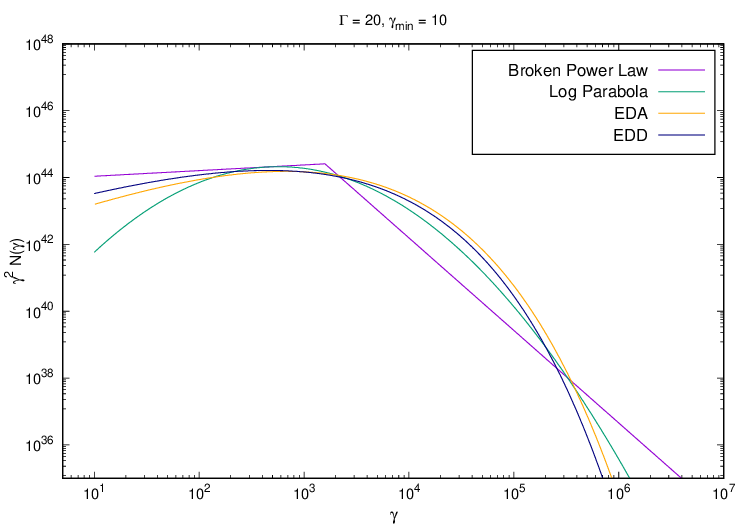}
    \caption{Particle energy distributions of PKS 1441+25 (left panel), Ton 599 (right panel) at $\Gamma = 20, T = 10^4, R = 10^{17}$, $\gamma_\text{min} = 10$ for all the models.}
    \label{fig: pd_pks_ton}
\end{figure*}

To illustrate the reason for no significant difference in jet power among the particle distributions considered for the FSRQs, we can look at the particle distributions shown in Figure \ref{fig: pd_pks_ton}. As noted by \cite{10.1093/mnras/stae706}, the $\gamma^2 N(\gamma)$ varied greatly across distributions, causing large differences in jet power for Broken Power-Law (BPL) models. In contrast, for FSRQs, the $\gamma^2 N(\gamma)$ values are nearly the same across the particle distributions considered, suggesting little to no notable difference in jet power in this study.

The calculated proton power component which are reported in Table \ref{gammamin=10_pks} and \ref{gammamin=10_ton} are less than that of HBL like Mkn 501. This can shed light on the fact that, in the case of the FSRQs, the proton power didn't seem to dominate, which might cause the jet power to be not sensitive to $\gamma_\text{min}$ variation.
 
\section*{Acknowledgements}
We acknowledge the use of public data from the NuSTAR, \textit{Swift}-XRT/UVOT from NASA’s High Energy Astrophysics Science Archive Research Center (HEASARC), \textit{Fermi}-LAT data from \textit{NuSTAR} Science Support Center (FSSC) of Goddard Space Flight Center. R.K. acknowledges the financial support of the NWU PDRF Fund NW.1G01487, the South African Research Chairs Initiative (grant No. 64789) of the Department of Science and Innovation and the National Research Foundation of South Africa. H.B. and R.G. would like to acknowledge IUCAA for their support and hospitality through their associateship program.

\section*{DATA AVAILABILITY}
The data and software used in this study can be accessed via NASA's HEASARC webpages and the Fermi Science Support Center (FSSC), for which the corresponding links and references are provided within the manuscript.

\bibliography{sample631}
\bibliographystyle{aasjournal}

\end{document}